\documentclass{article}

\usepackage{amsmath,amssymb}
\usepackage{graphicx}
\usepackage[font=small]{caption}
\usepackage{authblk}
\usepackage{hyperref}

\graphicspath{{./}{fig/}{}}

\renewenvironment{abstract}
{\small
 \begin{center}
 \bfseries \abstractname\vspace{-.5em}\vspace{0pt}
 \end{center}
 \list{}{%
  \setlength{\leftmargin}{3cm}%
  \setlength{\rightmargin}{\leftmargin}%
 }%
 \item\relax}
{\endlist}

\begin{document}

\title{Interval Spacing}
\author{Greg Kreider}
\affil{\small gkreider@primachvis.com}
\affil{\small Primordial Machine Vision Systems, Inc.}
\date{}

\maketitle

\begin{abstract}
\leftskip=5cm \rightskip=5cm
We define interval spacing as the difference in the order statistics of data
over a gap of some width.  We derive its density, expected value, and
variance for uniform, exponential, and logistic variates.  We show that
interval spacing is equivalent to running a rectangular low-pass filter over
the spacing, which simplifies the expressions for the expected values and
introduces correlations between overlapping intervals.
\\
\leftskip=0pt \rightskip=0pt
\end{abstract}

The theory behind spacing, the difference between consecutive order statistics,
is well-developed by now.  Occasionally one sees the difference over larger
gaps being used.  \cite{venter67} estimates the location of the mode as the
midpoint of the interval with the most data points, where the size of the
interval depends on the polynomial order of the data around the mode.
\cite{duembgen08} collects a test statistic over a range of gaps to
determine if data has a non-zero slope.  \cite{jammal89} and \cite{mirkh13}
study such test statistics in general for non-overlapping segments.

Let us call ``interval spacing'' the difference in order statistics over
distances greater than one.  Extending the notation used in \cite{pyke65}
with $ i $ the upper index and $ w $ the width of the interval such that
$ w < i \leq n $, the interval spacing $ D_{i,w} $ is
\begin{equation} \label{eq:diw}
D_{i,w} = T_{i} - T_{i-w}
\end{equation}
where $ T_{i} $ is the $ i^{\rm th} $ order statistic.  Alternative notations
that have been used are a center point $ j = i-(w/2) $ and radius
$ r_{n} = w/2 $ \cite[(2.1)]{venter67}, or start index $ j = i-w+1 $ and
endpoint $ k = i $ of the range \cite{duembgen08}.  In this notation the
spacing $ D_{i} = D_{i,1} $.

The density of the interval spacing
\begin{align} \label{eq:fdiw.gen}
f_{D_{i,w}}(y) & =
\begin{aligned}[t]
  S_{1} \int_{-\infty}^{\infty} &
  \left\{ F_{x}(x) \right\}^{i-w-1} \left\{ F_{x}(x+y) - F_{x}(x) \right\}^{w-1}
  \left\{ 1 - F_{x}(x+y) \right\}^{n-i} \\
  & \times f_{x}(x) ~ f_{x}(x+y) ~ dx
\end{aligned} \\
S_{1} & = \frac{n!}{(i-w-1)! ~ (w-1)! ~ (n-i)!} \nonumber
\end{align}
follows from the joint density of two order statistics
\cite[(8) and (31) with $ r = i-w $ and $ r'=i $]{wilks48}, where
$ F_{x}(x) $ and $ f_{x}(x) $ are distribution and density functions.
Expanding \cite[(2.4)]{pyke65} gives the same density, using $ k_1 = i-w $,
$ t_1 = x $, $ k_2 = i $, $ t_2 = x+y $, $ k_3 = n+1 $, $ t_3 = \infty $,
and $ t_0 = -\infty $.  \eqref{eq:fdiw.gen} simplifies to the spacing's
density function \cite[(2.7)]{pyke65} when $ w = 1 $; notably, the factor of
the distribution function raised to $ w-1 $ disappears.  The expected value
and variance follow normally from the first two moments of this density.
\begin{align} \label{eq:ediw.gen}
E\Bigl\{ D_{i,w} \Bigr\}
 & = \int_{0}^{\infty} y ~ f_{D_{i,w}}(y) ~ dy
     \displaybreak[0] \\
\label{eq:vdiw.gen}
V\Bigl\{ D_{i,w} \Bigr\}
 & = E\Bigl\{ D_{i,w}^2 \Bigr\} - E\Bigl\{ D_{i,w} \Bigr\}^2 
   = \int_{0}^{\infty} y^{2} ~ f_{D_{i,w}}(y) ~ dy - E\Bigl\{ D_{i,w} \Bigr\}^2
\end{align}

For uniform variates over the range $ a, b $ we have
\begin{align}
\label{eq:fdiw.unif}
f_{D_{i,w,unif}}(y)
 & = \frac{n!}{(w-1)! (n-w)!}
     ~ \left(\frac{1}{b-a}\right)^{n} y^{w-1} ~ (b-y-a)^{n-w}
     \displaybreak[0] \\
\label{eq:ediw.unif}
E\Bigl\{ D_{i,w,unif} \Bigr\}
 & = w ~ \frac{b-a}{n+1}
     \displaybreak[0] \\
 \label{eq:vdiw.unif}
V\Bigl\{ D_{i,w,unif} \Bigr\}
 & = w ~ \frac{(n+1-w)}{n+2} \left( \frac{b-a}{n+1} \right)^{2}
\end{align}

For exponential variates with rate parameter $ \lambda $ these are
\begin{align}
\label{eq:fdiw.exp}
f_{D_{i,w,exp}}(y)
 & = w ~ \binom{n-i+w}{n-i} ~ \lambda ~
     \left\{ e^{-\lambda y} \right\}^{n-i+1}
     \left\{ 1 - e^{-\lambda y} \right\}^{w-1}
     \displaybreak[0] \\
\label{eq:ediw.exp}
E\Bigl\{ D_{i,w,exp} \Bigr\}
 & = w ~ \binom{n-i+w}{n-i} ~ \frac{1}{\lambda} ~ (-1)^{w-1}
     \sum_{k=0}^{w-1} \binom{w-1}{k} \frac{(-1)^{k}}{(n-i+w-k)^{2}}
     \displaybreak[0] \\
\label{eq:ediwsq.exp}
E\Bigl\{ D_{i,w,exp}^{2} \Bigr\}
 & = w ~ \binom{n-i+w}{n-i} ~ \frac{2}{\lambda^{2}} ~
   \sum_{k=0}^{w-1} \binom{w-1}{k} \frac{(-1)^{k}}{(n-i+w-k)^{3}}
\end{align}
Use \eqref{eq:vdiw.gen} and the last two equations to calculate the variance;
squaring \eqref{eq:ediw.exp} and subtracting from \eqref{eq:ediwsq.exp} does
not lead to a simpler equation.  The pre-factor can also be written
$ (n-i+w)! / (w-1)! (n-i)! $, which will cancel a factor $ (w-1)! $ from
the combinatorial inside the series.

For logistic variates with mean $ \mu $ and standard deviation $ \sigma $
we find
\begin{align}
\label{eq:fdiw.logis}
f_{D_{i,w,logis}}(y)
 & = \frac{1}{\sigma} ~ S_{2} ~
     e^{y/\sigma} \left\{ e^{y/\sigma}-1 \right\}^{w-1} \nonumber \\
 & \qquad \times \begin{aligned}    
     {}_{2}F_{1}\left(i,n-i+w+1 ; n+w+1 ; 1-e^{y/\sigma} \right)
   \end{aligned} \\
S_{2}
 & = \frac{n!}{(n+w)!} ~ \frac{(n-i+w)!}{(n-i)!} ~ \frac{(i-1)!}{(i-w-1)!}
     ~ \frac{1}{(w-1)!} \nonumber \\
 & = w \prod_{j=1}^{w} \frac{(i-j) (n-i+j)}{j (n+j)}
     \nonumber \displaybreak[0]
\end{align}
\begin{align}
\label{eq:ediw.logis}
E\Bigl\{ D_{i,w,logis} \Bigr\}
 & = \sum_{k=0}^{w-1}
     \frac{n!}{(i-w-1)! (w-1-k)! (n-i+1+k)!} ~ \sigma ~ (-1)^{w-1-k}
     \nonumber \\
 & \qquad \times \left[
   \begin{aligned}
     \frac{1}{i-k-1} \left[ \psi(i-k) - \psi(i-k-1) \right] \\
     - \sum_{l=1}^{n-i+w-1} \frac{1}{n-i+w-l} ~ B(n-i+w+1-l, i-k-1) \\
     - ~ \frac{w-1-k}{n-i+w-1} ~ B(n-i+w+1, i-k-2) \\
     + \sum_{l=2}^{w-1-k} (-1)^{l}
            \sum_{j=0}^{l-1} S_{3} ~ B(n-i+w+1+j, i-k-2-j)
   \end{aligned} \right] \\
S_{3}
 & = \frac{(w-1-k)!}{(w-1-k-l)!} ~ \frac{(n-i+w-l-1)!}{(n-i+w-l+j)!}
     ~ \frac{1}{l (l-1-j)!}
   \nonumber
\end{align}
The second moment would require an additional integration by parts of these
terms, which we do not perform.  $ B(a,b) = (a-1)! (b-1)! / (a+b-1)! $ is the
beta function and $ \psi(x) $ the dilogarithm.

The interval spacing introduces an extra factor $ F^{w-1} $ of the
distribution function which requires an integration by parts to handle.  This
leads to more complicated expressions for the expected interval spacing than
for the normal spacing, although all results reduce to the spacing versions
found in \cite{kreider23a} if $ w = 1 $.  The uniform and exponential
equations follow from known definite integrals (see Supplemental Materials
for derivations).  Rather than integrating the hypergeometric function in
the logistic density function \eqref{eq:fdiw.logis} to get the first
moment, it is better to integrate first over $ y $ after combining
\eqref{eq:fdiw.gen} and \eqref{eq:ediw.gen}, then over $ x $.  This must be
done by parts, which involves a recursion down $ n $ that gives a series
that must then be integrated term by term.

High-precision math libraries must be used to evaluate the series.  The
factorial scaling factors in \eqref{eq:fdiw.unif} -- \eqref{eq:fdiw.logis}
reach $ n^{w} / w! $ while the spacing is of order one, so the series sum
of terms of alternating sign of this size must nearly cancel.

\begin{figure}[t]
\centering
\includegraphics[width=\textwidth]{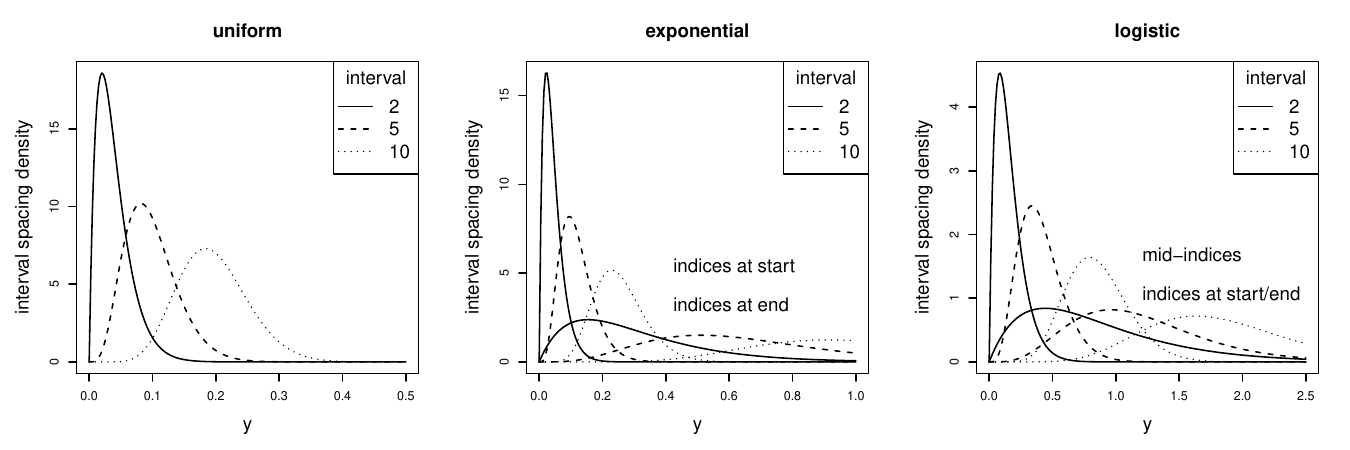}
\caption{\label{fig:fdiw} Density of the interval spacing at widths $ w $
of 2, 5, and 10 and indices $ i $ as noted.}
\end{figure}

Figure~\ref{fig:fdiw} plots the density of the interval spacing for draws of
$ n = 50 $ data points.  The function for uniform variates in the left graph
is independent of the index $ i $, but the others show two sets of curves.
The densities with high, narrow peaks correspond to small values from the
variate, found at the start of the exponential's order statistics, drawn
with $ i = w + 2 $, or the center of the logistic, drawn with $ i = n / 2 $.
The lower, broader densities come from the tails of the distribution where
the order statistics are largest.  This occurs near the final indices of
the expontial, drawn for $ i = n - 2 $, and the initial indices of the
logistic, drawn for $ i = w + 2 $ but also valid by symmetry at the other
tail.  All densities skew to the right, although this is much more noticeable
in the broader curves.

\begin{figure}
\centering
\includegraphics[width=\textwidth]{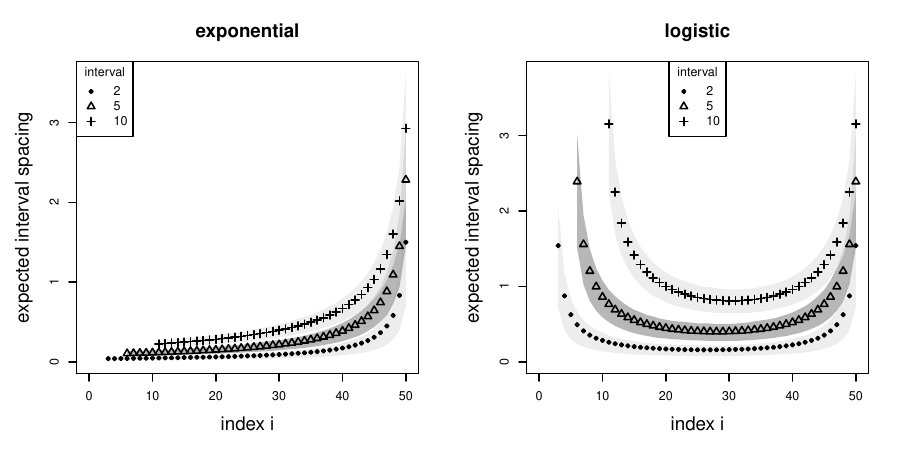}
\caption{\label{fig:eldiw} Expected interval spacing for exponential and
logistic variates at widths $ w $ of 2, 5, and 10.  Bands are inter-quartile
ranges from simulations.}
\end{figure}

Figure~\ref{fig:eldiw} plots the expected interval spacing for exponential
and logistic variates at three widths; the data size $ n = 50 $ has been kept
small for visibility.  The grey bands are inter-quartile ranges based on
ten thousand simulated draws.  The expected values are not centered, lying
closer to the upper quartile.  This reflects the skewing of the density.
Such shifting also happens for the medians, not plotted.

An interval can be broken into non-overlapping segments whose sub-spacing
will add.  For example, if $ w $ is even we can split the interval in half,
\begin{align} \label{eq:halfw}
D_{i,w/2} + D_{i-w/2,w/2}
  & = T_{i} - T_{i-w/2} + T_{i-w/2} - T_{i-w}
    = T_{i} - T_{i-w} \nonumber \\
  & = D_{i,w}
\end{align}
The half intervals sum to the whole.  Any sub-intervals need not have the
same size, which would happen in this example if $ w $ were odd, but the
endpoints must match and cover the whole range.  Splitting into more pieces
is also possible, a process which ultimately ends by breaking the interval
into $ w $ single steps.  The interval spacing is the sum of these spacings,
\begin{equation} \label{eq:disum}
D_{i,w} = \sum_{j=0}^{w-1} D_{i-j}
\end{equation}
A sum over consecutive data points is just a rectangular low-pass filter.
If the total were divided by $ w $ this would be a running mean.  Because
such normalization is not done --- the filter kernel is a vector of
$ w $ 1's rather than $ 1/w $'s --- larger intervals amplify
differences.  This must be taken into account when comparing different
widths.  The spacing is not constant over the interval so the interval
spacing is not just the spacing scaled by $ w $, although it is close,
especially where the variate's density is nearly constant, at the start
of one-sided distributions or the center of two-sided.

\begin{figure}
\centering
\begin{minipage}[t]{0.45\textwidth}
\includegraphics[width=\textwidth]{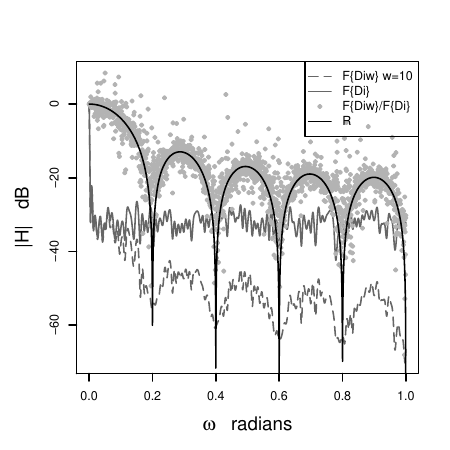}
\caption{\label{fig:lprect} Interval spacing is equivalent to a rectangular
low-pass filter applied to spacing.}
\end{minipage}
\qquad
\begin{minipage}[t]{0.45\textwidth}
\includegraphics[width=\textwidth]{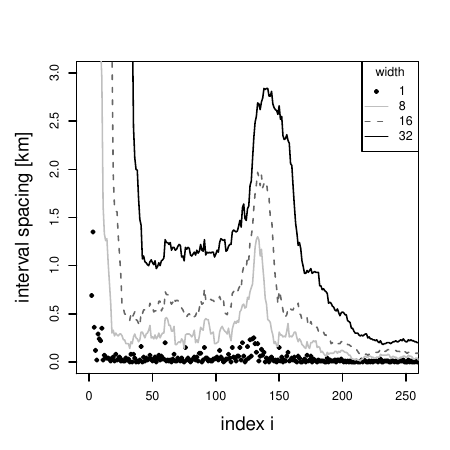}
\caption{\label{fig:lowpass} Low-pass filtering of spacing for increasing
interval widths $ w $.}
\end{minipage}
\end{figure}

We can demonstrate that filtering occurs by taking the ratio of the Fourier
transforms of $ D_{i,w} $ and $ D_{i} $ and comparing it to the impulse
response $ \mathcal{R} $ of the rectangular kernel $ r $.  That is,
\begin{equation} \label{eq:rectconv}
\mathcal{F} \left\{ D_{iw} \right\}
  = \mathcal{F} \left\{ r \ast D_{i} \right\}
  = \mathcal{R} \cdot \mathcal{F} \left\{ D_{i} \right\}
\end{equation}
Figure~\ref{fig:lprect} shows the ratio applied to a draw of 1200 points from
a uniform distribution.  The transforms of the spacing and interval spacing at
$ w = 10 $ have been smoothed for display, but their ratio has not.
Superimposed is the magnitude of $ \mathcal{R} $, which has the same number
of side lobes at the same height.

\eqref{eq:disum} also applies to the expected values, which allows us to
simplify \eqref{eq:ediw.exp} using $ E\bigl\{ D_{i,exp} \bigr\}
$ \cite[(8)]{kreider23a}
\begin{equation} \label{eq:ediw.exp.simple}
E\Bigl\{ D_{i,w,exp} \Bigr\}
 = \sum_{j=0}^{w-1} E\Bigl\{ D_{i,exp} \Bigr\}
 = \sum_{j=0}^{w-1} \frac{1}{\lambda ~ (n-i+j+1)}
\end{equation}
and \eqref{eq:ediw.logis} with $ E\bigl\{ D_{i,logis} \bigr\} $
\cite[(17)]{kreider23a}
\begin{equation} \label{eq:ediw.logis.simple}
E\Bigl\{ D_{i,w,logis} \Bigr\}
 = \sum_{j=0}^{w-1} \frac{\sigma ~ n}{(i - j - 1) (n - i + j + 1)}
\end{equation}
The expected spacing for uniform variates, $ (b - a) / (n + 1) $, is
repeated $ w $ times in the sum independently of $ j $, matching
\eqref{eq:ediw.unif}.  We could write an expression for the expected
interval spacing for Gumbel variates using \cite[(20)]{kreider23a}, although
the result is neither simple nor illuminating.  \eqref{eq:disum} can also be
used for variates requiring numeric integration for their expected spacing.

Unlike \eqref{eq:ediw.exp} and \eqref{eq:ediw.logis},
\eqref{eq:ediw.exp.simple} and \eqref{eq:ediw.logis.simple} do not require
high precision libraries and can be evaluated directly.

Although a rectangular low-pass filter has a wide main band, it suppresses
the sidelobes by only a moderate amount, which allows high-frequency residuals
to remain, especially at sharp edges.  Figure~\ref{fig:lowpass} plots the
spacing for the depth of earthquakes under Mt. St. Helens before its eruption
in 1980 \cite{scott92}.  Depths are considered below the surface and are
negative, so the first order statistics are the deepest and the largest
interval spacings are at the smallest indices and initially decrease rapidly.
There are five individual large spacings between indices 40 and 100
corresponding to a depth of -7.90~--~-5.62~km, and a cluster around 130,
between -4.83~km and -2.87~km.  The former create small coarse bumps in the
$ w = 8 $ interval spacing, and the latter a sharp peak.  This peak widens
with the interval width, while the bumps merge into a single rough region
without damping their range, an example of insufficient suppression of higher
frequencies.  Eventually at $ w = 32 $ one individual point falls within
every interval and the bumps disappear.  The signal simplifies to a nearly
constant level, becoming a plateau or flat.  The trailing edge of the peak
beyond index 150 does become smoother with larger intervals.

\begin{figure}
\centering
\begin{minipage}[t]{0.45\textwidth}
\includegraphics[width=\textwidth]{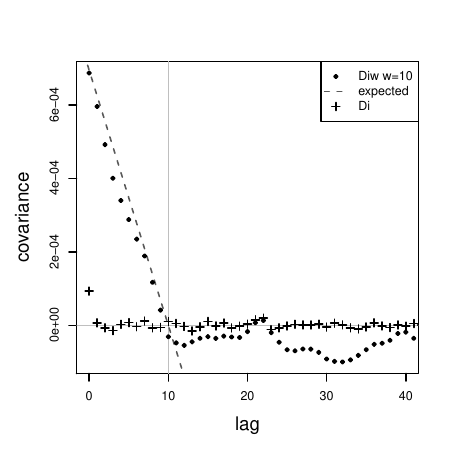}
\caption{\label{fig:cov} Auto-covariance of the interval spacing follows the
self-convolution of the rectangular kernel.}
\end{minipage}
\end{figure}

If we consider overlapping intervals by lag $ 1 \le l < w $,
\begin{align} \label{eq:diwdiff}
D_{i,w} - D_{i-l,w}
  & = \sum_{k=0}^{w-1} D_{i-k} - \sum_{k=0}^{w-1} D_{i-l-k}
    = \sum_{k=0}^{w-1} D_{i-k} - \sum_{k'=0}^{w-1+l} D_{i-k'} \nonumber \\
  & = \sum_{k=0}^{l-1} D_{i-k} - \sum_{k'=w}^{w-1+l} D_{i-k'}
\end{align}
where in the second step we have shifted $ k' = l + k $.  The two terms in
the final result do not overlap, because the common spacings
$ \sum_{k=l}^{w-1} D_{i-k} $ have canceled.  Said differently, overlapping
intervals will be correlated, sharing $ w-l $ terms.  The autocovariance of
the interval spacing follows the convolution of the rectangular kernel (not
its impulse response) with itself, which is a linear decrease with initial
value equal to the variance of $ D_{i,w} $ and slope
$ - V\left\{ D_{i,w} \right\} / w $ (Figure~\ref{fig:cov}).

\section*{Supplemental Material}
The supplement for this paper includes derivations of the density, expected
value, and variance of the interval spacing for uniform, exponential, and
logistic variates.

\bibliographystyle{amsplain}
\bibliography{dmodal}

\end{document}